%% file: AFTgroupLT_2024-01-05.tex
\documentclass[12pt]{article}
\usepackage{amsmath}
\usepackage{graphicx,psfrag,epsf}
\usepackage{enumerate}
\usepackage{url} 

\usepackage{natbib}

\usepackage{bm}
\usepackage{amsmath,amssymb}
\usepackage{graphicx}
\usepackage{color}
\usepackage{mathtools}
\usepackage{setspace}
\usepackage{float} 

\usepackage{hyperref}

\input Macros

\newcommand{\blind}{0}

\mathtoolsset{showonlyrefs=true} 

\begin{document}

\def\spacingset#1{\renewcommand{\baselinestretch}%
{#1}\small\normalsize} \spacingset{1}


\if0\blind
{
  \title{\bf Group lasso priors for Bayesian accelerated failure time models with left-truncated and interval-censored data}
  \author{Harrison T. Reeder\hspace{.2cm}\\
	Massachusetts General Hospital, Boston, Massachusetts, U.S.A. \\ 
	Harvard Medical School, Boston, Massachusetts, U.S.A. \\ 
	\\
Sebastien Haneuse and Kyu Ha Lee\\
	Harvard T.H. Chan School of Public Health, Boston, Massachusetts, U.S.A.}
	\date{}
  \maketitle
} \fi

\if1\blind
{
  \bigskip
  \bigskip
  \bigskip
  \begin{center}
    {\LARGE\bf Title}
\end{center}
  \medskip
} \fi

\bigskip
\begin{abstract}
An important task in health research is to characterize time-to-event outcomes such as disease onset or mortality in terms of a potentially high-dimensional set of risk factors. For example, prospective cohort studies of Alzheimer's disease typically enroll older adults for observation over several decades to assess the long-term impact of genetic and other factors on cognitive decline and mortality. The accelerated failure time model is particularly well-suited to such studies, structuring covariate effects as `horizontal' changes to the survival quantiles that conceptually reflect shifts in the outcome distribution due to lifelong exposures. However, this modeling task is complicated by the enrollment of adults at differing ages, and intermittent followup visits leading to interval censored outcome information. Moreover, genetic and clinical risk factors are not only high-dimensional, but characterized by underlying grouping structure, such as by function or gene location. Such grouped high-dimensional covariates require shrinkage methods that directly acknowledge this structure to facilitate variable selection and estimation. In this paper, we address these considerations directly by proposing a Bayesian accelerated failure time model with a group-structured lasso penalty, designed for left-truncated and interval-censored time-to-event data. We develop an accessible R package with a custom Markov chain Monte Carlo sampler for efficient estimation, and investigate the impact of various methods of penalty tuning and thresholding for variable selection. We present a simulation study examining the performance of this method relative to models with an ordinary lasso penalty, and apply the proposed method to identify groups of predictive genetic and clinical risk factors for Alzheimer's disease in the Religious Orders Study and Memory and Aging Project (ROSMAP) prospective cohort studies of AD and dementia.
\end{abstract}


\spacingset{1.45} 

\section{Introduction}

Alzheimer's disease (AD) is a degenerative neurologic condition that can arise among older adults, leading to progressive cognitive decline and memory loss, dementia and death. An urgent scientific goal is to identify and characterize factors associated with increased risk of developing AD, as well as those that may be protective against AD onset. Towards this goal, typical prospective cohort studies of Alzheimer's disease enroll older adults for periodic assessment over a period of decades to assess the long-term impact of genetic and other factors on cognitive decline and mortality. Examples of such studies in the US include the Adult Changes in Thought (ACT) study \citep{kukull2002dementia}, the Baltimore Longitudinal Study of Aging \citep{kawas2000agespecific}, and the Religious Orders Study and Memory and Aging Project (ROSMAP) \citep{bennett2018religious}. More generally, similar designs are used for numerous large-scale `lifecourse epidemiology' studies of longitudinal health outcomes, such as the Nurse's Health Study \citep{baer2011risk} and the Growing Up Today study \citep{corliss2008sexual}. 
However, the design of such studies poses unique challenges to the analysis of time-to-event outcomes such as AD onset or mortality. 

The most common time-to-event outcome analysis method used in this context is the Cox proportional hazards model, which structures covariate effects as constant `hazard ratios' across time. Not only can these ratios be difficult to interpret \citep{hernan2010hazards,uno2015alternatives}, but it is often implausible to consider a constant hazard ratio (i.e., `proportional hazards') over long timescales such as commonly seen in cohort studies of AD \citep{bellera2010variables}. By contrast, the accelerated failure time (AFT) model is particularly well-suited to such studies, structuring covariate effects as interpretable multiplicative changes on the survival quantiles, and conceptually characterizing underlying shifts in the outcome distribution due to lifelong exposures \citep{reeder2023characterizing}. Regardless of model structure, prospective cohort studies of AD also enroll patients at varying advanced ages, leading to potential for selection bias due to left truncation (or `delayed entry') that must be accounted for in the modeling approach \citep{guo1993eventhistory}. And finally, because of the long timescale follow-up visits are intermittent, and exact event timing between visits is typically unknown, leading to information loss known as `interval censoring' that must be acknowledged in the analysis \citep{lee2017accelerated}.  

In the frequentist literature, a variety of penalized estimation methods have been developed for statistical modeling to identify important covariates from a high-dimensional set of candidates, including the lasso and other sparsity-inducing penalties \citep{tibshirani1996regression,fan2001variable}. Of particular interest are penalties that encode available information about the structure of the covariates---the fused lasso penalizes differences between coefficients, while the group lasso enforces sparsity on pre-specified groups of covariates \citep{tibshirani2005sparsity,yuan2006model}. Common examples of `grouped' covariates are indicator variables for levels of a categorical variable, but can more generally encompass any \emph{a priori} grouping information, such as genetic variables sharing a gene location, or clinical variables measuring a common function. 

Tools for structured variable selection and estimation have also been developed the Bayesian paradigm \citep{raman2009bayesian,kyung2010penalized}. In particular, the Bayesian group lasso prior retains the benefits of the frequentist group lasso penalty in incorporating known information about covariate relationships, but final variable selection is not restricted to inclusion or exclusion of entire covariate blocks as in the frequentist setting. This is especially relevant in the context of risk modeling for complex diseases like AD, as there are many important potential clinically-measured predictors of risk. These predictors naturally form groups based on function and collinearity, such as complementary measures of physical activity, or pain in different parts of the body. However, not every measure may be prognostic within each group.

Recent advances have expanded the literature on Bayesian variable selection for time-to-event analyses, but important gaps remain particularly in light of the challenges arising in the study of illnesses such as AD with complex etiologies unfolding over long timescales. Several Bayesian AFT models with various sparse priors or spike-and-slab variable selection schemes have been proposed, which rely on Markov chain Monte Carlo (MCMC) sampling schemes that cannot readily extend to left-truncated and interval-censored data \citep{konrath2015bayesian,ahmad2017clinically,lee2017variable,newcombe2017weibull,chang2018scalable,zhang2018bayesian,zhu2019bayesian}. Likewise, several frequentist penalized Cox models have been developed to accommodate either left-truncation and/or interval-censoring, but cannot readily incorporate an AFT covariate effect specification \citep{li2020adaptive,zhao2020simultaneous,du2021unified,mcgough2021penalized}. 
\citet{huang2019bilevel} 
consider a frequentist AFT model with a ``group exponential lasso penalty'' to encode structured sparsity, however their approach also does not accommodate left truncation or interval censoring. To our knowledge, there have not been high-dimensional AFT models developed in either paradigm that also admit left truncation and interval censoring, let alone that can further incorporate covariate grouping information into the specification.

To address this gap, we propose a Bayesian approach to high-dimensional AFT modeling with a group lasso shrinkage prior on the covariates, and an efficient custom MCMC sampling scheme to accommodate left truncation and interval censoring. This approach applies data augmentation to resolve all forms of complex censoring following the work of 
\citet{tanner1987calculation}. 
Using the resulting complete-data likelihood corrected for left-truncation we develop an efficient custom Gibbs sampler to draw from the posterior using a combination of Gibbs, Metropolis-Hastings, and Hamiltonian Monte Carlo (HMC) updates \citep{neal2011mcmc}. We also use an empirical Bayes approach for the group lasso regularization parameter via Monte Carlo Expectation Maximization (MCEM) to update the parameter during sampling, and develop a thresholding method for variable selection from the posterior samples. These methods are explored via simulation studies, and illustrated with application to the task of risk modeling for onset of AD or dementia using data from the Religious Orders Study and Memory and Aging Project (ROSMAP) prospective cohort studies of AD and dementia \citep{bennett2018religious}. All methods are implemented in the publicly available R package \texttt{psbcGroup} on CRAN.

\section{Methods}\label{sec:method}
\subsection{The Accelerated Failure Time Model}

\begin{figure}[ht]
\begin{center}
	\minipage{0.5\textwidth}%
	\includegraphics[width=\linewidth]{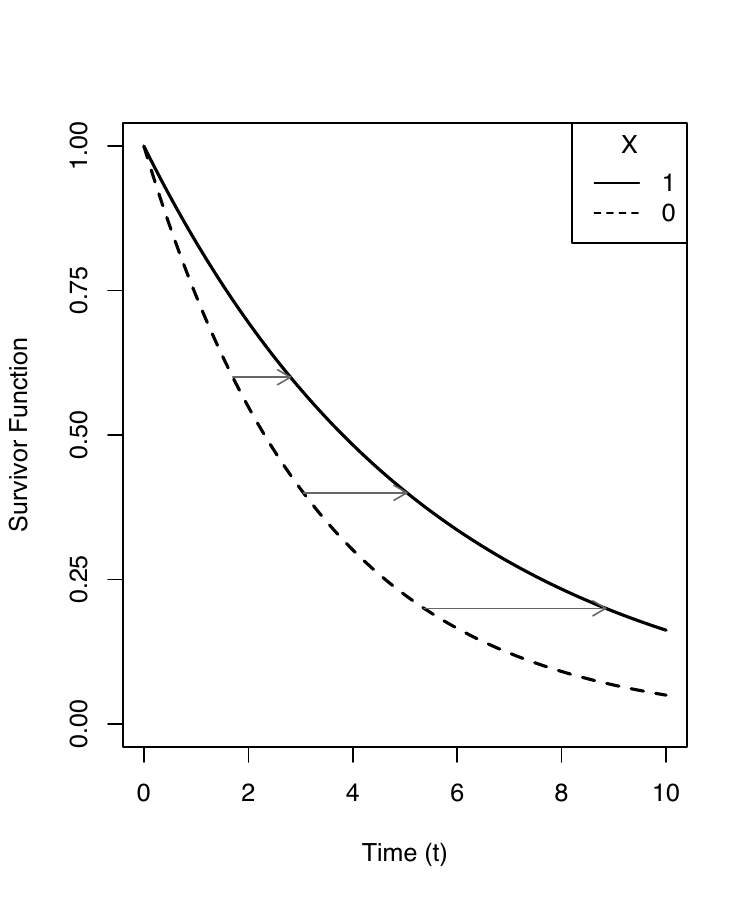}
	\endminipage\hfill
	\minipage{0.5\textwidth}
	\includegraphics[width=\linewidth]{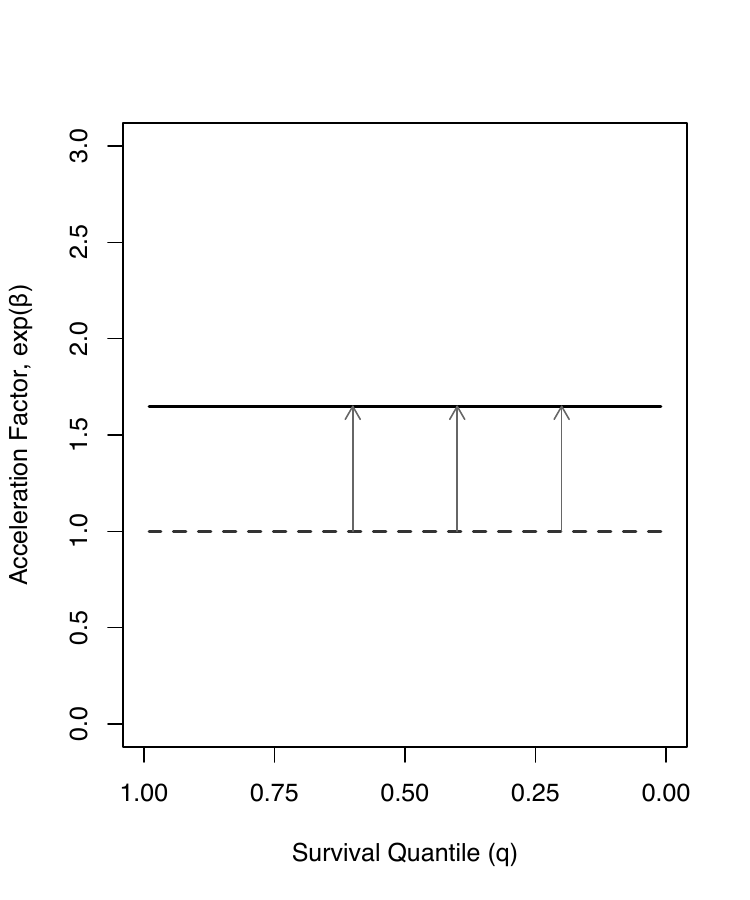}
	\endminipage
	\caption{Sample survivor curves (left panel) and corresponding multiplicative acceleration factors across survival quantiles (right panel). Baseline survivor function shown is $S_0(t) = \exp(-0.3t)$, with $\beta=0.5$.\label{fig:SAF1}}
\end{center}
\end{figure}

Suppose interest is in characterizing a time-to-event outcome $T$, such as time of AD onset, in terms of a $p$-dimensional set of risk factors $\bfX$ (allowing $p>n$) subject to shrinkage and selection, and an additional low-dimensional set of $q$ clinical covariates denoted $\bfZ$ not subject to shrinkage. This distinction reflects that in practical applications the analyst may not want to apply shrinkage to all covariates. The AFT model structures this relationship via a log-linear model of time,
\begin{equation}
	\log(T) =  \bfX^{\top}\bfbeta + \bfZ^{\top}\bfgamma + \epsilon, \label{AFTmodel}
\end{equation}
where $\epsilon$ is a random error term and $\bfbeta$ and $\bfgamma$ are corresponding vectors of regression coefficients. To understand the role of $\epsilon$, note that exponentiating this term yields a positive random variable $T_0 = \exp(\epsilon)$, which follows the `baseline' time-to-event distribution corresponding to $\bfX=\boldsymbol 0$ and $\bfZ=\boldsymbol 0$. If this baseline survival distribution has survivor function $S_0(t) = \Pr(T_{0}\leq t)$, then the survivor function for $T$ when $\bfX=\bfx$ and $\bfZ=\bfz$ follows from \eqref{AFTmodel} as
\begin{equation}
	S(t\mid \bfx, \bfz) = S_0(t \times \exp(-\bfx^{\top}\bfbeta - \bfz^{\top}\bfgamma)). \label{AFTmodel2}
\end{equation}
Based on this structure, interpreting the regression coefficients follows by considering any survival quantile of interest $0 < q <1$. For example, median survival ($q=0.5$) is the time by which half of the population will have experienced the event. Then the $q$th quantile survival time for $x_j + 1$ is multiplied by a factor of $\exp(\beta_j)$ relative to $x_j$ (holding other covariates constant). This relationship is constant across all values of $q$, and therefore the exponentiated coefficient $\exp(\beta)$ is referred to as an `acceleration factor' which directly describes the multiplicative shift of the entire event time distribution. For a hypothetical univariable AFT model, Figure~\ref{fig:SAF1} illustrates this shift on the scale of the survivor function, and correspondingly the constant multiplicative effect across quantiles.

\subsection{Observed Data and Likelihood}
The time-to-event outcome of the $i$th subject is represented by the random variable $T_i$, corresponding to the subject's covariates $\bfX_i, \bfZ_i$ and having density $f(t_i \mid \bfX_i, \bfZ_i)$. However, by the design of the study the subject may have delayed entry---that is, rather than enrolling at the time origin of 0, they have a study entry time represented by $C_{0,i}$. By definition, $T_i > C_{0,i}$ among those who enter the study, which has potential to induce selection bias if not acknowledged. The density of the event time conditional on study entry is the event time density inverse weighted by the probability of having survived to the time of study entry: 
\begin{equation}
f(t_i \mid c_{0,i}, \bfx_i, \bfz_i) = \frac{f(t_i \mid \bfx_i, \bfz_{i})}{\int_{c_{0,i}}^{\infty} f(s \mid \bfx_i, \bfz_{i})ds}, \label{condpdf}
\end{equation}
which would represent subject $i$'s likelihood contribution under the hypothetical `complete' data denoted $\mathcal{D}^*_i = \{c_{0,i}, \bfx_i, \bfz_i, t_i \}$. However, when study visits are intermittent it may only be known that the event occurred between two visit times, $c_{L,i} \leq T_i \leq c_{U,i}$, which encompasses standard right-censoring when $c_{U,i}=\infty$. Then we denote the actual observed data $\mathcal{D}_i = \{c_{0,i}, \bfx_i, \bfz_i, c_{L,i}, c_{U,i}\}$, and the missing event time $t_i$ is framed as an unknown latent parameter on the interval $[c_{L,i}, c_{U,i}]$. As described below, in the Bayesian paradigm these latent event times can be sampled using a data augmentation scheme, enabling modeling via the likelihood \eqref{condpdf} computed on the augmented data.

\subsection{Bayesian Specification and the Group Lasso Prior} \label{sec:prior}

Towards the goal of model fitting in the presence of high dimensions and structured sparsity, we place a group lasso prior on the regression coefficients $\bfbeta$. Suppose that covariates can be partitioned \emph{a priori} into groups $k=1,\dots, K$ each of size $m_k \geq 1$, denoted $\bfX = (\bfX_{1}^{\top}, \dots,\bfX_{K}^{\top})^{\top}$. We will analogously denote each $n\times m_k$ design submatrix as $\mathbb{X}_k$, and the complete design matrix $\mathbb{X} = (\mathbb{X}_1,\dots,\mathbb{X}_K)$. This may arise with grouped indicator variables corresponding to a categorical covariate, or clustered covariates. Similarly partitioning the coefficients $\bfbeta=(\bfbeta_{1}^{\top}, \dots, \bfbeta_{K}^{\top})^{\top}$, then the group lasso prior is
\begin{equation}
\pi(\bfbeta|\sigma^2)\propto\exp\left(-\frac{\lambda}{\sqrt{\sigma^2}}\sum_{k=1}^K\|\bfbeta_k\|_{G_k}\right), \label{condPriorGL}
\end{equation}
where each $G_k$ is a pre-specified positive definite weight matrix, and $\|\bfbeta\|_{G_k} = (\bfbeta^{\top}G_k\bfbeta)^{1/2}$. 
A common choice is to weight each group by $m_k^{1/2}$, corresponding to $G_k = m_k\cdot I$, which will be our focus.

\citet{kyung2010penalized} 
have shown that for each $\bfbeta_k$ independently, the group lasso prior can be rewritten hierarchically as a multivariate normal prior with variance scaled by a gamma-distributed hyperparameter $\tau_k$:
\begin{align}\label{hierarchPriorGL}
    \bfbeta_k|\sigma^2, \tau_k^2 &\sim \textrm{MVN}\left(\bfzero_{m_k}, \sigma^2\tau^2_k G_{k}\right)
    \\ \pi(\tau_k^2|\lambda^2) &\propto \frac{\lambda^2}{2}(\tau_k^2)^{(m_k-1)/2}\exp\left(-\frac{ m_k\lambda^2\tau_k^2}{2}\right)
\end{align}
where $\bfzero_{m_k}$ is a zero vector of length $m_k$ and $\lambda^2$ is a global regularization hyperparameter. 

As a practical matter, shrinkage estimation typically requires covariates be standardized to have zero mean and unit variance, which in the group lasso context may leave substantial collinearity between intra-group variables. Instead, we follow the approach of 
\citet{simon2012standardization} 
by orthonormalizing variables by group before model fitting. Namely, for each $\mathbb{X}_k$, we compute the QR-decomposition $\mathbb{X}_k = \mathbf{Q}_k\mathbf{R}_k$, where $\mathbf{Q}_k$ is an $n\times m_k$ matrix with orthonormal columns, and $\mathbf{R}_k$ is an $m_k \times m_k$ upper triangular matrix. Model fitting is done using the group-orthonormalized covariate matrix $\mathbf{Q} = (\mathbf{Q}_1^{\top},\dots,\mathbf{Q}_K^{\top})$. Then using the relationship $\mathbb{X}_k\bfbeta_{k} = \mathbf{Q}_k(\mathbf{R}_k\bfbeta_{k})$, estimates are back-transformed to the original scale of the covariates by left-multiplying the fitted coefficients for each group by $\mathbf{R}_k^{-1}$. This has been shown to hold attractive theoretical properties in the frequentist setting, and better aligns with the use of the group lasso for general cluster-correlated covariates, as occurs in many applications. When covariates are in groups of size one (i.e., the `ordinary' lasso prior), this also reduces to typical variable-wise standardization. Again, variables and corresponding regression coefficients are back-transformed to their original scale after model fitting for interpretability.

For the additional low-dimensional set of `clinical covariates' $\bfZ$ not subject to shrinkage, we adopt independent normal priors $\gamma_j \sim N(0,v^2)$, $j=1,\dots,q$, sharing a common prior variance $v^2$.

Finally, we must specify a prior distribution for $\epsilon$, which in turn corresponds to the baseline survival distribution. Well-specified parametric priors on the error term have several advantages in this setting: they are computationally efficient, yield well-defined survival distributions across all quantiles, and can lead to improved efficiency in smaller samples. Therefore, we specify a normal distribution $\epsilon_i \sim \textrm{Normal}(\mu, \sigma^2)$, 
which induces a log-Normal survival distribution for $T_i$ having density
\begin{equation}
 	f(t_i|\bfx_i, \bfz_{i};\bfbeta, \bfgamma, \mu, \sigma^2) = \frac{1}{\sigma t_i}\phi\left(\frac{\log(t_i)-\mu-\bfx_i^{\top}\bfbeta - \bfz_{i}^{\top}\bfgamma}{\sigma}\right). \label{lh:para}
\end{equation}	
The $\mu$ parameter in \eqref{lh:para} can be interpreted as an additional intercept in the log-linear model \eqref{AFTmodel}, and $\sigma^2$ is a scale parameter governing the shape of the outcome distribution. Notice that this is the same $\sigma^2$ incorporated into the group lasso prior above, following 
Park and Casella \citep{park2008bayesian}. 
We place a normal prior on this intercept $\mu \sim \textrm{Normal}(\mu_0,h_0)$ and an inverse-Gamma prior on the scale parameter $\sigma^2 \sim \textrm{inverse-Gamma}(a_{\sigma}, b_{\sigma})$, where $\mu_0$, $h_0$, $a_{\sigma}$, and $b_{\sigma}$ are specified by the analyst.

\subsection{Bayesian Computation}

Because of its log-linear structure, the AFT model naturally lends itself to survival modeling in the Bayesian paradigm, because data augmentation can be used to resolve right censoring and even interval censoring, often yielding a straightforward linear model. Having fully specified a distribution for the outcome, the idea introduced by 
\citet{tanner1987calculation} 
and detailed in Appendix A is to treat each censored $t_i \in [c_{L,i}, c_{U,i}]$ as a latent parameter, and sample realizations from the truncated distribution. This augmentation yields `complete' outcome data, for which the standard AFT model reduces to the well-studied task of Bayesian linear regression. As most shrinkage priors have been developed initially for linear regression, the resulting sampling schemes follow directly \citep{lee2017variable}.

A key challenge addressed by this paper is that the above approach cannot be applied in the presence of left-truncation; the augmented complete data likelihood \eqref{condpdf} no longer has a standard form, and therefore does not reduce to a typical linear regression model. In particular this means that updates to the regression parameters $\bfbeta$ and $\bfgamma$ can no longer leverage conjugacy, and instead a custom sampling scheme must be developed.

Details on each update step are provided in Appendix A of the supplementary materials. In summary, we propose a novel Gibbs sampler combining data augmentation of the interval-censored event times, Gibbs updates for $\bftau$, Metropolis-Hastings updates for $\sigma^2$, and HMC updates to efficiently sample the (possibly high-dimensional) regression parameters $\bfbeta$ and $\bfgamma$. Finally,  at each iteration we implemented an empirical Bayes step for $\lambda^2$ using an MCEM approach as proposed by 
Park and Casella \citep{park2008bayesian}, 
which updates the parameter by maximizing a Monte Carlo estimate of the expected value of the log-likelihood under the $\lambda^2$ from the previous iteration.

\subsection{Variable Selection via SNC-BIC Thresholding Procedure}

Using the group lasso prior defined in \eqref{condPriorGL} induces shrinkage in the regression parameter posterior, but does not itself yield sparse estimates. Therefore, in this section we present a method for variable selection based on posterior samples.

In frequentist penalized estimation, variable selection typically proceeds by tuning a regularization parameter to optimize a model fit metric, such the Bayesian Information Criterion (BIC). Criteria like the BIC approximate overall model fit based on parameter point estimates, which excludes the information on model uncertainty available following Bayesian posterior sampling.

Instead, in the Bayesian paradigm 
Li and Lin \citep{li2010bayesian} 
propose a parameter-wise metric of posterior concentration around zero called the scaled neighborhood criterion (SNC), defined for each $\beta_j$ as $\text{SNC}_j = \Pr\left(|\beta_j| > \sqrt{\Var(\beta_j\mid \mathcal{D})}\right)$. In turn, they propose a model selection routine by first selecting all variables with SNC above a threshold $\kappa$, and reporting a final model including only the selected variables. However, the choice of $\kappa$ is itself unclear, and must be selected via some procedure.

Following 
Lee et al. \citep{lee2017variable}, 
we propose a model selection procedure called SNC-BIC, which combines elements of these techniques to select a sparse final model while acknowledging estimate uncertainty. First, using the posterior samples $\widetilde{\boldsymbol\beta}$ from the full model, identify a set of candidate models according to a grid of SNC thresholds $\kappa_m=m/M$ for $m = 1,\dots,M$. The number of candidate models might be less than $M$, as some $\kappa_m$ values might select the same set of variables, so $M=1000$ in our analyses. Let $\mathit{S}_m = \{j : \text{SNC}_j > \kappa_m\}$ be the set of coefficients selected for threshold $m$, and define the fitted linear predictor values for the $m$th candidate model and $i$th individual as $w_{m,i} = \bfz_{i}^{\top}\bfgamma + \sum_{j \in \mathit{S}_m} x_{i,j} \beta_j$. The number of selected covariates for candidate model $m$ is denoted $p_m$. Second, for each candidate model $m$ we fit a univariable frequentist log-normal AFT model by maximum likelihood, with $\mathbf{w}_{m}$ as a single unpenalized predictor variable. The maximized log-likelihood value is denoted $L_m$, and represents a relative metric of in-sample predictive performance for candidate model $m$. This methodology is similar to the debiasing step proposed by 
Tibshirani \citep{tibshirani2009univariate} 
for estimating a rescaling factor following shrinkage estimation. Third, a BIC-type information criterion is constructed using these components of the form BIC$_m = L_m - p_m\log n$. The final selected model is the minimizer of BIC$_m$, and the final estimates are the corresponding selected elements of $\widetilde{\boldsymbol\beta}$.  Accessible code to run the above sampler, and the proposed variable thresholding method, are available in the R package \texttt{psbcGroup} on CRAN.

\section{Simulation Studies} \label{sec:simulation}

In this section we present simulation studies evaluating the variable selection and estimation performance of the proposed AFT model with group lasso prior across a range of settings. The primary purpose of these simulations is to compare the group lasso prior and ordinary lasso prior in settings with high-dimensional covariates where grouping is present, and to assess how performance varies with possible misspecification of the baseline survival distribution.

\subsection{Simulation Setup}

For 100 simulation iterations, we generate data on $n=500$ individuals having $p=1000$ high-dimensional covariates $\bfX$, and $q=1$ additional confounding variable $Z$. To simulate grouped covariate data, we generate four latent values $V_k\sim U(0.8k - 2.2, 0.8k - 0.2)$ for $k=1,\dots,4$, from which we generated blocks of five clustered covariates centered around each latent value: $X_{j} \sim V_k + e_j$ where $e_j \sim Normal(0,0.01)$ for $j= 5k-4, 5k-3,\dots, 5k$. 
The remaining covariates $X_{j}$ for $j=21,\dots,1000$, and the  confounding variable $Z$, were independently drawn from a standard normal.

A sparse set of 20 truly non-zero covariate effects was specified with varying signal strength and to overlap with the grouped covariates: $\bfbeta = (\bfbeta^{*\top},-\bfbeta^{*\top},\bfbeta^{*\top},-\bfbeta^{*\top}, \bfzero_{p-20}^{\top})^{\top}$ where $\bfbeta^* = (-2, -1.5, -1, 1.5, 2)^{\top}$. The confounding variable effect was set at $\gamma=-1$.

Finally, outcomes were generated under the AFT model structure \eqref{AFTmodel}, with four different specifications for the distribution of $\epsilon$: a log-Normal model matching the parametric AFT specification, as well as three more complex alternative distributions described in Table \ref{tab:simsetting}. Outcomes were administratively right-censored to achieve an observed event rate of 65-75\% across simulation settings. Also, while the specified baseline distributions have dramatically different shapes, they all have similar mean and variance, to facilitate comparison of results across varying forms of misspecification.  Plots illustrating the form of each baseline distribution are shown in Appendix B of the supplementary materials.

\begin{table}[ht]
\centering
\caption{Simulation scenarios}\label{tab:simsetting}
\begin{tabular}{ll}
  \hline
 & Distribution of $\epsilon$ \\ 
  \hline
Scenario 1 (log-Normal) & Normal(4, 1) \\ 
  Scenario 2 (bimodal) & Mixture of Normal(2.8, 0.01) and Normal(5.2, 0.01) \\ 
  Scenario 3 (heavy-tailed) & Mixture of Normal(4, 2) and Normal(4, 0.01) \\ 
    Scenario 4 (right-skewed) & Skew-normal(2.8, 1.7, 20)\citep{azzalini2013skew} \\ 
   \hline
\end{tabular}
\end{table}

For each simulation, we fit group lasso and ordinary lasso comparator models using a log-Normal baseline survival specification, setting uninformative priors via the hyperparameters $\mu_{0}=0$ and $h_0=10^6$ for $\mu$, and $a_{\sigma}=b_{\sigma}=0.7$ for $\sigma$. Two chains of 50,000 MCMC iterations are run, with 50\% burn-in. For the group lasso prior, groups were set to match the true covariate groupings, with independent lasso priors on the remaining covariates. Final models were selected using the SNC-BIC thresholding procedure.

For each model in each simulation setting, we report the mean $\ell_2$ estimation error $\| \widetilde{\bfbeta}-\bfbeta\|_2$, and the mean and standard deviation of several variable selection metrics:  true positive rate (proportion of truly non-zero $\beta$'s selected); false positive rate (proportion of truly zero $\beta$'s selected); positive predictive value (proportion of selected $\beta$'s that is truly non-zero); and negative predictive value (proportion of non-selected $\beta$'s that is truly zero). We also report the mean number of regression parameters selected.

\subsection{Results}

Mean estimation error results are presented in Table~\ref{tab:simerror}. Overall, the estimation error for each comparator model was consistent across the simulation scenarios, indicating that regression parameter estimation is relatively robust to misspecification of the baseline distribution. The relative weakest performance for both comparators occurred in Scenario 2, which has a bimodal distribution of $\epsilon$. This is a uniquely challenging setting, and may have been disproportionately distorted by administrative censoring affecting only the righthand mode of the distribution. 

Across all scenarios there is consistently less estimation error when using the group lasso prior, relative to the ordinary lasso. This performance gain corresponds to the additional information contained in the group specification, relative to the ordinary lasso which ignores the high correlation of covariates within each group. 

\begin{table}[ht]
\centering
\caption{Mean and standard deviation of $l_2$ estimation error of $\bfbeta$ across simulations.}\label{tab:simerror}
\begin{tabular}{lcc}
  \hline
Setting & Group Lasso & Ordinary Lasso \\ 
  \hline
Scenario 1 & 3.12 (0.44) & 5.45 (0.38)  \\ 
  Scenario 2 & 3.42 (0.58) & 5.77 (0.43) \\ 
  Scenario 3 & 2.99 (0.46) & 5.37 (0.43) \\ 
    Scenario 4 & 2.94 (0.43) & 5.36 (0.38) \\ 
   \hline
\end{tabular}
\end{table}

Mean selection performance metrics are presented in Table~\ref{tab:simselect}. As with estimation error, the variable selection performance remained consistent across the scenarios, again with some underperformance in Scenario 2 evident. Taken together, these metrics indicate that the SNC-BIC method generally selected highly sparse models, with average number of selected parameters ranging from 7-11 depending on method and scenario. The false positive rate was near zero across all scenarios, indicating that truly zero parameters were almost never selected, while the true positive rates show that a substantial proportion of truly nonzero parameters are nonetheless excluded from the final model.

Importantly, all selection performance metrics are consistently higher for models fit with a group lasso prior compared to the ordinary lasso prior. Again, this captures the performance impact of leveraging the covariate grouping information in the prior, relative to independently penalizing the covariates. 

\begin{table}[ht]
\begin{center}
\caption{Mean and standard deviation of selection performance metrics of $\bfbeta$ across simulations.}\label{tab:simselect}
\begin{tabular}{lllllll}
  \hline
Setting & Model & TPR & FPR & PPV &  NPV & Selected \\ 
  \hline
  Scenario 1 & Group Lasso & 50.8 (5.1) & 0.0 (0.1) & 96.4 (6.2) & 99.0 (0.1)& 10.5 (1.6) \\
                    & Ordinary Lasso & 32.6 (11.8) & 0.1 (0.1) & 92.1 (11.9) & 98.6 (0.2) & 7.2 (2.8) \\
  Scenario 2 & Group Lasso & 39.1 (8.8) & 0.0 (0.0) & 98.3 (4.2) & 98.6 (0.2) & 8.1 (1.9) \\
                    & Ordinary Lasso & 14.0 (9.3) & 0.0 (0.1) & 82.2 (30.3) & 90.4 (26.8)& 3.3 (2.0) \\
  Scenario 3 & Group Lasso & 51.3 (4.5) & 0.1 (0.1) & 96.8 (5.7) & 99.1 (0.1)& 10.8 (1.3) \\
                    & Ordinary Lasso & 38.4 (15.3) & 0.1 (0.1) & 90.6 (14.0) & 97.8 (9.9)& 8.5 (3.6) \\
  Scenario 4 & Group Lasso & 51.5 (5.1) & 0.1 (0.1) & 95.6 (7.6) & 98.9 (0.2)& 10.8 (2.0) \\
                    & Ordinary Lasso & 38.1 (13.5) & 0.1 (0.2) & 89.8 (12.3) & 98.8 (0.3)& 8.7 (3.4) \\
   \hline
\end{tabular}
\end{center}
Abbreviations: true positive rate (TPR); false positive rate (FPR); positive predictive value (PPV); negative predictive value (NPV); group lasso (GL); ordinary lasso (OL)
\end{table}

%
%
%

\section{Data Application}

To illustrate the proposed AFT model, we consider the task of modeling clinical risk factors for AD or dementia onset in older adults using data collected by the ongoing ROSMAP cohort studies, which began respectively in 1994 and 1997 \citep{bennett2018religious}. This setting features a long timescale, delayed entry, interval censored outcomes, and a large number of covariates that naturally group into clinical domains.

In this analysis, we identified {498} subjects enrolled {without AD or dementia between ages 65 and 86}, who were followed either until withdrawal or death and who had complete data on {40} clinical risk factors assessed at study enrollment. The study design includes annual cognitive screening to diagnose onset of AD or dementia, with death status monitored continuously. The outcome of AD or dementia diagnosis is interval censored in these one year intervals, and death is treated as a censoring mechanism to yield a cause-specific analysis. The timescale is age with a time origin of 65, and subjects who enroll after age 65 are subject to left truncation (or ``delayed entry'') based on their age at enrollment. The median age of study entry was {78.4}. Over the followup period, {84 (16.8\%)} of participants had an observed AD or dementia diagnosis, with a median age at diagnosis of {88.4}.

The clinical risk factors under study are summarized in Table~\ref{tab:demo}, and for the group lasso prior, we formed these risk factors into {20} groups based on clinical function. {We again set uninformative priors via the hyperparameters} $\mu_{0}=0$ and $h_0=10^6$ for $\mu$, and $a_{\sigma}=b_{\sigma}=0.7$ for $\sigma$. We ran two chains of 50,000 samples with 50\% burn in, and used SNC-BIC to select a final model under both the group lasso prior and ordinary lasso prior.

\begin{table}[ht]
\begin{center}
\caption{Participant characteristics at study entry, by observed outcome status.}\label{tab:demo}
\resizebox{\textwidth}{!}{\begin{tabular}{llccc}
  \hline
 & & Total  & Observed AD/Dementia &  Withdrawal/Death before \\
Clinical Grouping & Characteristic& (n=498) & Diagnosis (n=84) &  AD/Dementia Diagnosis (n=414) \\ 
  \hline
genetic & APOE-$\epsilon$4 Genetic Variant & 92 (18.5\%) & 23 (27.4\%) & 69 (16.7\%) \\ 
  genetic & APOE-$\epsilon$3 + TOMM40 Genetic Variant & 86 (17.3\%) & 15 (17.9\%) & 71 (17.1\%) \\ 
  sex & Male Sex & 131 (26.3\%) & 20 (23.8\%) & 111 (26.8\%) \\ 
  racial\_ethnic & Hispanic Ethnicity & 25 (5.0\%) & 6 (7.1\%) & 19 (4.6\%) \\ 
  racial\_ethnic & White Self-Reported Race & 461 (92.6\%) & 78 (92.9\%) & 383 (92.5\%) \\ 
  marital & Married at Study Entry & 467 (93.8\%) & 81 (96.4\%) & 386 (93.2\%) \\ 
  education & 15+ Years of Education & 265 (53.2\%) & 34 (40.5\%) & 231 (55.8\%) \\ 
  education & Maternal 15+ Years of Education & 60 (12.0\%) & 7 (8.3\%) & 53 (12.8\%) \\ 
  education & Paternal 15+ Years of Education & 88 (17.7\%) & 7 (8.3\%) & 81 (19.6\%) \\ 
  ses & Low Income at Study Entry & 91 (18.3\%) & 15 (17.9\%) & 76 (18.4\%) \\ 
  ses & Low Income at Age 40 & 136 (27.3\%) & 35 (41.7\%) & 101 (24.4\%) \\ 
  substance\_use & Ever Smoking Status & 218 (43.8\%) & 27 (32.1\%) & 191 (46.1\%) \\ 
  substance\_use & Less than one Alcoholic Drink per Month (Lifetime) & 217 (43.6\%) & 43 (51.2\%) & 174 (42.0\%) \\ 
  cancer & History of Cancer & 171 (34.3\%) & 29 (34.5\%) & 142 (34.3\%) \\ 
  blood\_pressure & History of Hypertension & 289 (58.0\%) & 45 (53.6\%) & 244 (58.9\%) \\ 
  stroke & History of Stroke & 34 (6.8\%) & 11 (13.1\%) & 23 (5.6\%) \\ 
  diabetes & History of Diabetes & 71 (14.3\%) & 12 (14.3\%) & 59 (14.3\%) \\ 
  heart\_conditions & History of Heart Disease & 42 (8.4\%) & 6 (7.1\%) & 36 (8.7\%) \\ 
  blood & Anemia at Study Entry & 64 (12.9\%) & 13 (15.5\%) & 51 (12.3\%) \\ 
  blood & MCV (fL) at Study Entry & 92.402 (5.131) & 92.875 (4.148) & 92.307 (5.307) \\ 
  cardiac\_rx & Lipid-lowering Medication Use at Study Entry & 216 (43.4\%) & 33 (39.3\%) & 183 (44.2\%) \\ 
  bmi & BMI 18.5-25 at Study Entry & 154 (30.9\%) & 30 (35.7\%) & 124 (30.0\%) \\ 
  bmi & BMI 25-30 at Study Entry & 189 (38.0\%) & 30 (35.7\%) & 159 (38.4\%) \\ 
  bmi & BMI 30-35 at Study Entry & 101 (20.3\%) & 16 (19.0\%) & 85 (20.5\%) \\ 
  bmi & BMI 35+ at Study Entry & 50 (10.0\%) & 8 (9.5\%) & 42 (10.1\%) \\ 
  motor & Dexterity Score at Study Entry & 1.044 (0.132) & 1.003 (0.143) & 1.053 (0.129) \\ 
  motor & Gait Score at Study Entry & 1.051 (0.239) & 1.004 (0.237) & 1.060 (0.238) \\ 
  motor & Hand Strength Score at Study Entry & 1.069 (0.288) & 0.952 (0.263) & 1.093 (0.288) \\ 
  physical & Hours of Physical Activity at Study Entry & 3.555 (3.854) & 3.441 (4.302) & 3.578 (3.762) \\ 
  physical & Independent Activities of Daily Living Score at Study Entry & 0.564 (0.909) & 0.702 (1.050) & 0.536 (0.876) \\ 
  physical & Basic Activities of Daily Living (Katz) Score at Study Entry & 0.082 (0.385) & 0.095 (0.399) & 0.080 (0.382) \\ 
  physical & Mobility (Rosow-Breslau) Score at Study Entry & 0.586 (0.882) & 0.512 (0.857) & 0.601 (0.887) \\ 
  joint\_pain & Hand Pain at Study Entry & 100 (20.1\%) & 15 (17.9\%) & 85 (20.5\%) \\ 
  joint\_pain & Foot Pain at Study Entry & 69 (13.9\%) & 13 (15.5\%) & 56 (13.5\%) \\ 
  joint\_pain & Hip Pain at Study Entry & 101 (20.3\%) & 15 (17.9\%) & 86 (20.8\%) \\ 
  joint\_pain & Knee Pain at Study Entry & 127 (25.5\%) & 19 (22.6\%) & 108 (26.1\%) \\ 
  neck\_back\_pain & Lower Back Pain at Study Entry & 207 (41.6\%) & 30 (35.7\%) & 177 (42.8\%) \\ 
  neck\_back\_pain & Neck pain at Study Entry & 77 (15.5\%) & 8 (9.5\%) & 69 (16.7\%) \\ 
  mental\_health & Depressive Symptom (CES-D) Score at Study Entry & 0.932 (1.520) & 1.333 (2.026) & 0.850 (1.384) \\ 
  mental\_health & Mental Health Medication Use at Study Entry & 127 (25.5\%) & 25 (29.8\%) & 102 (24.6\%) \\ 
   \hline
\end{tabular}}
\end{center}
Note: binary variables are reported as `n (\%)' and continuous variables reported as `mean (standard deviation)'. 
Abbreviations: mean corpuscular volume (MCV); femtoliter (fL); Center for Epidemiological Studies Depression Scale (CES-D)
\end{table}

\begin{figure}[ht]
\begin{center}
	\includegraphics[width=\linewidth]{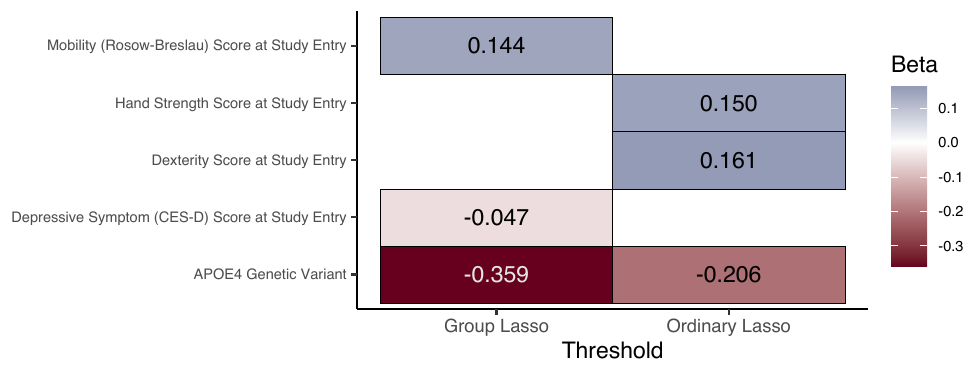}
\end{center}
\caption{Median posterior regression parameter estimates $\widetilde{\bfbeta}$ for selected coefficients \label{fig:fit_figure}}
\end{figure}

The results for each model are presented in Figure~\ref{fig:fit_figure}.
The group lasso AFT and ordinary lasso AFT each selected three covariates into the final models. Both models selected the well-known genetic risk factor APOE-$\epsilon$4 as the strongest risk factor, corroborating evidence found in previous research \citep{strittmatter1996apolipoprotein}. The group lasso, for example, estimated that having the APOE-$\epsilon$4 gene is associated with a median age at AD/dementia diagnosis that is $\exp(-0.15) = 0.861$ times lower than not having that variant.

The group lasso also selected a measure of increased depressive symptoms in older age as associated with earlier time to diagnosis \citep{wilson2002depressive}, and {the Rosow-Breslau score, a measure of decreased mobility, as associated with later time to diagnosis.} These associations were not selected by the ordinary lasso model however, which instead identified two metrics of improved motor function as having associations with longer time to diagnosis. While it is somewhat surprising that decreased mobility would seem to have a protective association with time to AD or dementia onset, this association was also seen in sensitivity analyses performed using a frequentist Cox model with lasso penalty, as reported in Appendix C of the supplementary materials. Additionally, the result is a cause-specific association treating death as a censoring mechanism, and may have a complex interplay with the association of mobility and mortality, and with other variables.

Overall, these results indicate modest associations between several clinical risk factors and time until subsequent diagnosis of AD or dementia, with the nature and magnitude of the findings depending on the choice of shrinkage prior. However, based on the results seen in simulations, the group lasso prior's handling of collinearity among grouped clinical risk factors representing similar functions may be more appropriate in this setting.

\section{Discussion}

Many cohort studies, particularly of older adults, are characterized by varied age at study entry and periodic follow-up visits, which naturally subject time-to-event outcomes of interest to left truncation and interval censoring. The proposed AFT model acknowledges both phenomena, and structures covariate effects as `acceleration factors' that naturally lend themselves to interpretation in long-term cohort follow-up, as multiplicative changes to survival quantiles such as median survival time.

Models with a group lasso prior consistently outperformed those that applied independent lasso priors when known covariate grouping information was available. In settings where grouping information is not available, analysts might first apply a clustering method to the covariates to infer group memberships, which can then be incorporated into the group lasso prior \citep{lee2017variable}. Simulations also indicated that the SNC-BIC thresholding method yields very sparse final models, with low false negative rates across all models and scenarios, and modest true positive rates that were consistently higher among the group lasso. Other thresholding methods may yield differing operating characteristics, such as screening covariates with credible intervals non-overlapping with zero, or using clustering methods to distinguish estimates into non-zero and zero groups \citep{li2017variable}. 

While models with parametric baseline specification were the focus of this paper to facilitate efficient computation, the simulations indicate that estimation error and selection performance was relatively robust even when the parametric baseline was misspecified. Nevertheless, extension to non-parametric baseline specifications, such as using a Dirichlet process mixture of Normals for $\epsilon$, could be undertaken at the cost of computational efficiency \citep{lee2017accelerated}.

A major contribution of the proposed model is the use of HMC sampling for the regression parameters, to achieve substantial efficiency gains in the presence correlated high-dimensional covariates. This approach simultaneously updates all regression parameters, using gradient information to efficiently explore the high-dimensional posterior geometry. In contrast, simpler random walk samplers with parameter-wise updates can be very slow to explore the posterior and slow to converge \citep{gelman2014bayesian}. This gain in per-iteration efficiency particularly benefits this model as each iteration requires Monte Carlo simulation to update the regularization parameter $\lambda^2$. Moreover, while use of HMC sampling methods has increased through the use of the Stan programming language, the proposed custom HMC sampler efficiently combines HMC with other Gibbs updates and data augmentation, which cannot otherwise be implemented in Stan \citep{carpenter2017stan}. The proposed method is efficiently coded in C, and wrapped in the accessible R package \texttt{psbcGroup} on CRAN.

\section*{Acknowledgments}

This project was supported by the National Institute of Child Health and Development [grant number F31HD102159 to HTR], the National Institute of Dental and Craniofacial Research [grant number R03DE027486 to KHL], and the National Institute of General Medical Sciences [grant number R01GM126257 to KHL]. The National Institutes on Aging supported the Religious Orders Study [grant numbers P30AG010161 and R01AG015819], the Rush Memory and Aging Project [grant number R01AG017917], and related pipelines [grant numbers U01AG046152 and U01AG061356].

We thank the study participants and staff of the Rush Alzheimer’s Disease Center. ROSMAP resources can be requested at \url{https://www.radc.rush.edu}.


\bibliographystyle{apalike}
\bibliography{sources}

\newpage

\appendix 

\makeatletter
\renewcommand{\thetable}{\thesection.\@arabic\c@table}
\@addtoreset{table}{section}
\makeatother

\makeatletter
\renewcommand{\thefigure}{\thesection.\@arabic\c@figure}
\@addtoreset{figure}{section}
\makeatother

\makeatletter
\renewcommand{\theequation}{\thesection.\@arabic\c@equation}
\@addtoreset{equation}{section}
\makeatother



\section*{Appendix Introduction} \label{sec:appintro}

In this appendix we present additional details and results beyond what could be presented in the main manuscript. To distinguish the two documents, alpha-numeric labels are used in this document while numeric labels are used in the main paper. Section~\ref{app:mcmcdetail} provides additional detail about the Bayesian Markov chain Monte Carlo sampling scheme. Section~\ref{app:simdet} provides additional details of the simulation studies. Section~\ref{app:dataapp} provides additional results from the data application.

\section{Sampler} \label{app:mcmcdetail}

Let $\Theta = \{\bfbeta, \bfgamma, \mu, \sigma^2, \bftau^2, \lambda^2\}$ and $\Theta^{-(\bfbeta)}$ denote a set of parameters $\Theta$ with $\bfbeta$ removed. 

We define the complete data likelihood correcting for left truncation as in \eqref{condpdf} of the main text, using the pdf for log-Normal outcome data given by \eqref{lh:para} in the main text.

Then multiplying by the priors for each component of $\Theta$ as outlined in Section~\ref{sec:prior} yields the posterior up to a normalizing constant. We then draw samples using a Gibbs sampling scheme by iterating through the following steps:
\begin{enumerate}
	\item Data augmentation
	
	For each $i$=1,\ldots,$n$, we impute/update ${t_i}$ by drawing from the log-Normal distribution on the truncated interval ($c_{ij}$, $c_{ij+1}$).
	
	\item Updating $\bfbeta$
	
	Let $k_{(j)}$ denote the $k$th group to which $j$th covariate belongs. Assuming $G_k=m_kI_k$ as in the text, the full conditional for $\bfbeta$ is given by
	\be
		\pi(\bfbeta | \Theta^{-(\bfbeta)}) \propto \prod_{i=1}^n\frac{f(t_i| \bfx_i, \bfz_{i}; \Theta)}{\int_{c_{0i}}^{\infty} f(s | \bfx_i, \bfz_{i}; \Theta)ds} \times \prod_{j=1}^p \exp\left(-\frac{\beta_j^2}{2\sigma^2\tau_{k_{(j)}}^2}\right).
	\ee
	Since this does not have a standard form, we use the Hamiltonian Monte Carlo (HMC) algorithm \citep{neal2011mcmc}. Letting $\eta_i = \mu + \bfx_i^{\top}\bfbeta+\bfz_i^{\top}\bfgamma$, the gradient, $\partial\log\pi(\bfbeta|\Theta^{-(\bfbeta)})/\partial\bfbeta^{\top}$, is given by $\bfX^{\top}\bfa + \bfb$, where $\bfa = (a_1,\ldots,a_n)^{\top}$ and $\bfb = (b_1,\ldots,b_p)^{\top}$ with
	\be
		a_i &=& \frac{\log(t_i) -\eta_i}{\sigma^2}-\frac{1}{\sigma}\phi\left(\frac{\eta_i-\log(c_{0i})}{\sigma}\right)\left\{\Phi\left(\frac{\eta_i-\log(c_{0i})}{\sigma}\right)\right\}^{-1}, \nonumber \\
		b_j &=& -\frac{\beta_j}{\sigma^2\tau_{k(j)}^2}.
	\ee
	
	\item Updating $\bfgamma$
	
	The full conditional for $\bfgamma$ is given by
	\be
		\pi(\bfgamma | \Theta^{-(\bfgamma)}) \propto \prod_{i=1}^n\frac{f(t_i| \bfx_i, \bfz_{i}; \Theta)}{\int_{c_{0i}}^{\infty} f(s | \bfx_i, \bfz_{i}; \Theta)ds} \times \prod_{j=1}^q \exp\left(-\frac{\gamma_j^2}{2v^2}\right).
	\ee
	
	Since this does not have a standard form, we use the HMC algorithm using the following gradient:
	\be
		\frac{\partial\log\pi(\bfgamma|\Theta^{-(\bfgamma)})}{\partial\bfgamma^{\top}}=\bfZ^{\top}\bfa-\bfc, \nonumber
	\ee
	where $\bfc=(c_1,\ldots,c_q)^{\top}$ with $c_j=-\gamma_j/v^2$.
	
	\item Updating $\sigma^2$
	
	The full conditional for $\sigma^2$ is given by
	
	\be
		\pi(\sigma^2| \Theta^{-(\sigma^2)}) \propto \prod_{i=1}^n\frac{f(t_i| \bfx_i, \bfz_{i}; \Theta)}{\int_{c_{0i}}^{\infty} f(s | \bfx_i, \bfz_{i}; \Theta)ds} \times \prod_{j=1}^p \exp\left(-\frac{\beta_j^2}{2\sigma^2\tau_{k(j)}^2}\right) \times \pi(\sigma^2),
	\ee
	where $\pi(\sigma^2)$ is the pdf of the Inverse-Gamma($a_{\sigma}$, $b_{\sigma}$) distribution. We use a Metropolis Hastings update based on Normal proposal distribution for $\log(\sigma^2)$, centered at the current value and with a pre-specified variance. 
		\item Updating $\mu$: The full conditional for $\mu$ is given by
	\be
		\pi(\mu | \cdot) \propto \prod_{i=1}^n\frac{f(t_i|\bfbeta, \bfbeta_c, \mu, \sigma^2, \bfx_i, \bfx_{c,i})}{\int_{c_{0i}}^{\infty} f(s | \bfbeta, \bfbeta_c, \mu, \sigma^2, \bfx_i, \bfx_{c,i})ds} \times \frac{1}{\sqrt{h_0}}\phi\left(\frac{\mu-\mu_0}{\sqrt{h_0}}\right).
	\ee
	We used a random walk MH step where we use a Normal proposal distribution centered at the current value of the parameter with pre-specified variance.

	\item Updating $\bftau^2$
	    
	The full conditional of $1/\tau^2_k$ is an inverse-Gaussian distribution given by
	\be
    		\frac{1}{\tau^2_k} | \cdot \sim \text{Inverse-Gaussian}\left(\sqrt{\frac{ m_k \lambda^2\sigma^2}{\|\bfbeta_k\|^2}}, ~ m_k\lambda^2\right).
	\ee
	
	\item Updating $\lambda^2$
	
	To estimate $\lambda^2$, we use Monte Carlo EM algorithm \citep{park2008bayesian}. The E-step in the $t$-th iteration of the algorithm involves taking the expected value of the log-likelihood conditional on $\mathcal{D}_i$ to calculate
	\be
		Q(\lambda^2 | \lambda^{2 (t-1))}) = \frac{p+K}{2}\log(\lambda^2) - \frac{\lambda^2}{2}\sum_{k=1}^K m_k E_{\lambda^{2 (t-1)}}\left[\tau_k^2|\mathcal{D}_i\right]+c, \nonumber 
	\ee 
	where $c$ includes terms that do not involve $\lambda^2$. In the M-step:
	\be
		\lambda^{2 (t)} = \frac{p+K}{\sum_{k=1}^{K} m_k E_{\lambda^{2 (t-1)}}\left[\tau_k^2|\mathcal{D}_i\right]}.
	\ee
	The expectation is calculated through a Monte Carlo average of the output from the Gibbs sampler.
	
\end{enumerate}

\newpage

\section{Additional Simulation Details} \label{app:simdet}

\begin{figure}[H]
\begin{center}
	\includegraphics[width=0.9\linewidth,page=1]{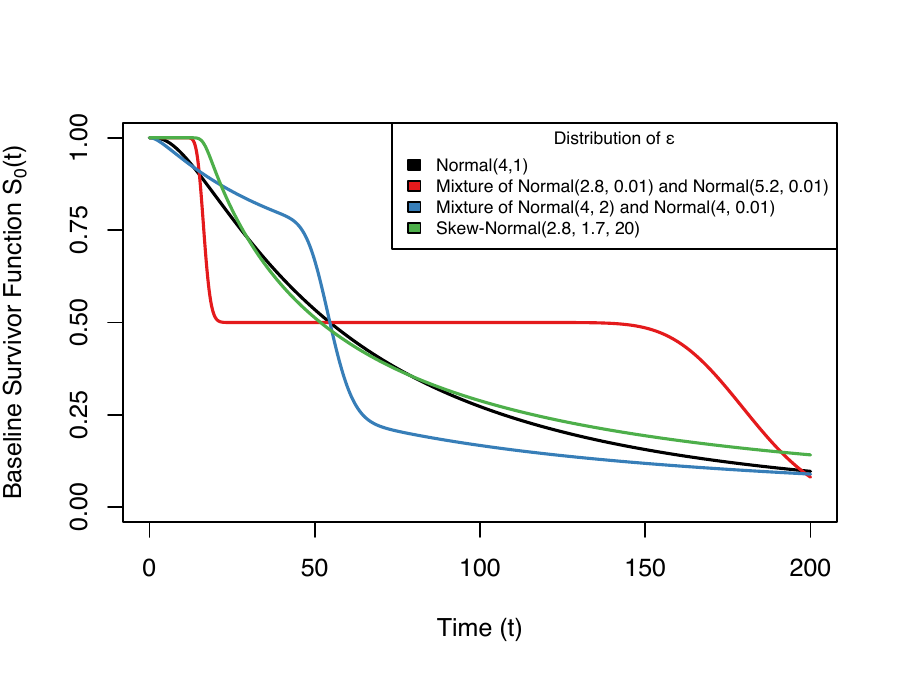}
\end{center}
\caption{Baseline survivor functions for distribution of $T_0$ across simulation settings. \label{fig:suppfigsim}}
\end{figure}

\newpage

\section{Additional Data Application Results} \label{app:dataapp}

\begin{figure}[H]

\begin{center}
	\includegraphics[width=0.7\linewidth,page=1]{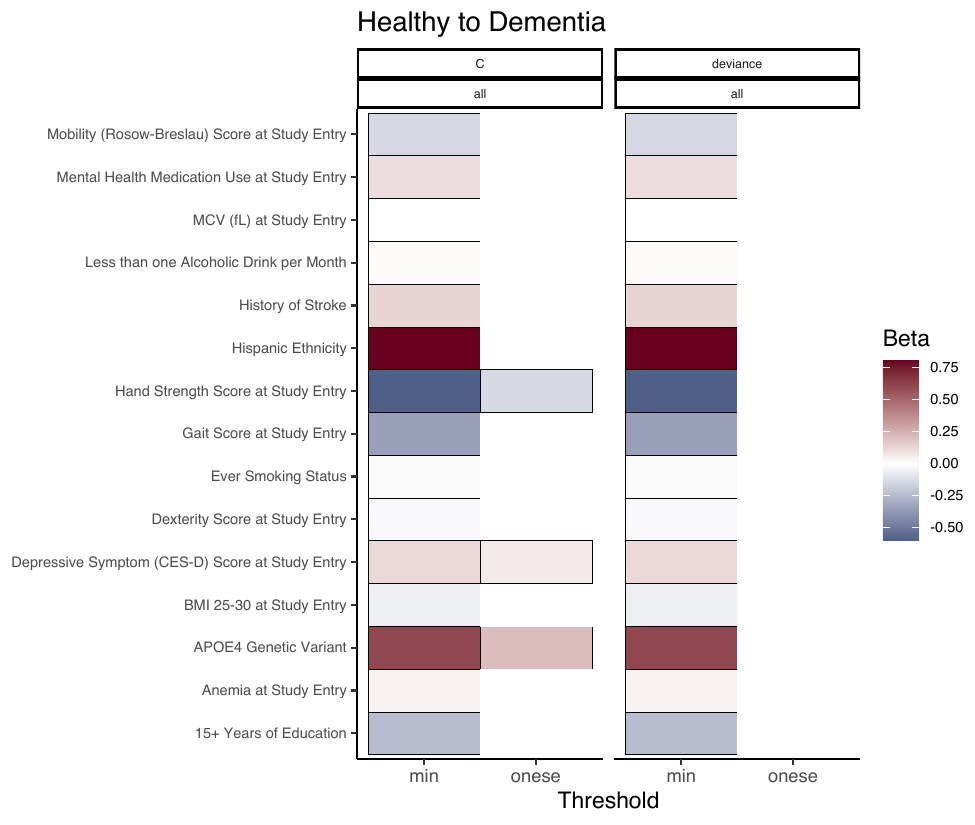}
\end{center}
\caption{Lasso-penalized Cox proportional hazards model coefficient estimates. Regularization parameter $\lambda$ tuned via 10-fold cross validation with respect to either C-index or partial likelihood (`deviance'), selecting either minimizer (`min') or largest $\lambda$ having cross-validated metric estimate within one standard deviation of the minimum (`onese'). \label{fig:suppfig}}
\end{figure}

As a sensitivity, we fit lasso-penalized frequentist Cox proportional hazards models to the same data as the data application, using midpoint imputation to resolve interval censoring, and accounting for left truncation in the partial likelihood \citep{mcgough2021penalized}. For comparison, the value of $\lambda$ was selected using cross-validation under four different rules implemented in the R package \texttt{glmnet}, and the results presented in Figure~\ref{fig:suppfig}. Specifically, for each of two performance metrics---the $C-index$ and the partial likelihood `deviance'---we selected $\lambda$ under one of two rules: the $\lambda$ value that minimizes the cross-validated metric (`min'), and alternatively the largest $\lambda$ that yields a cross-validated metric still within one standard deviation of the minimized value (`onese'). For more detail, see \texttt{glmnet} documentation.

\end{document}

%% file: Macros.tex
\newcommand{\nmathbf}{\bm}

\def\bfX{\nmathbf X}

\def\bfZ{\nmathbf Z}

\def\bfa{\nmathbf a}
\def\bfb{\nmathbf b}
\def\bfc{\nmathbf c}

\def\bfx{\nmathbf x}

\def\bfz{\nmathbf z}

\def\bfbeta   {\nmathbf \beta}
\def\bfgamma  {\nmathbf \gamma}

\def\bftau    {\nmathbf \tau}

\newcommand{\bfzero}{{\nmathbf 0}}

\def\boldfacefake#1{\kern-4pt
   \hbox{ \mathsurround=0pt
   \hbox to 0.4pt{$#1$\hss}\hbox to 0.4pt{$#1$\hss}\hbox {$#1$}}}

\newcommand{\Var}{\mbox{Var}}

\newcommand{\btable}{\begin{table}[h]\centering}
\newcommand{\etable}{\end{table}}
\newcommand{\bt}{\begin{parag}\small \let\b=\nsb \let\sb=\nssb \begin{tabular}}
\newcommand{\et}{\end{tabular}\let\b=\nb \let\sb=\nsb\end{parag}}

\newenvironment{parag}{\par}{\par}

\newcommand{\be}{\begin{eqnarray}}
\newcommand{\ee}{\end{eqnarray}}
\newcommand{\ba}{\begin{eqnarray*}}
\newcommand{\ea}{\end{eqnarray*}}

\newcommand{\reals}{\mbox{\rm I\kern-.20em R}}
\newcommand{\sreals}{\mbox{\small \rm I\kern-.20em R}}


%% file: AFTgroupLT_2024-01-05.bbl
\begin{thebibliography}{}

\bibitem[Ahmad and Fr{\"o}hlich, 2017]{ahmad2017clinically}
Ahmad, A. and Fr{\"o}hlich, H. (2017).
\newblock Towards clinically more relevant dissection of patient heterogeneity
  via survival-based {{Bayesian}} clustering.
\newblock {\em Bioinformatics}, 33(22):3558--3566.

\bibitem[Azzalini, 2013]{azzalini2013skew}
Azzalini, A. (2013).
\newblock {\em The skew-normal and related families}, volume~3.
\newblock Cambridge University Press.

\bibitem[Baer et~al., 2011]{baer2011risk}
Baer, H.~J., Glynn, R.~J., Hu, F.~B., Hankinson, S.~E., Willett, W.~C.,
  Colditz, G.~A., Stampfer, M., and Rosner, B. (2011).
\newblock Risk {{Factors}} for {{Mortality}} in the {{Nurses}}' {{Health
  Study}}: {{A Competing Risks Analysis}}.
\newblock {\em American Journal of Epidemiology}, 173(3):319--329.

\bibitem[Bellera et~al., 2010]{bellera2010variables}
Bellera, C.~A., MacGrogan, G., Debled, M., {de Lara}, C.~T., Brouste, V., and
  {Mathoulin-P{\'e}lissier}, S. (2010).
\newblock Variables with time-varying effects and the {{Cox}} model: {{Some}}
  statistical concepts illustrated with a prognostic factor study in breast
  cancer.
\newblock {\em BMC Medical Research Methodology}, 10(1):20.

\bibitem[Bennett et~al., 2018]{bennett2018religious}
Bennett, D.~A., Buchman, A.~S., Boyle, P.~A., Barnes, L.~L., Wilson, R.~S., and
  Schneider, J.~A. (2018).
\newblock Religious {{Orders Study}} and {{Rush Memory}} and {{Aging Project}}.
\newblock {\em Journal of Alzheimer's Disease}, 64(s1):S161--S189.

\bibitem[Carpenter et~al., 2017]{carpenter2017stan}
Carpenter, B., Gelman, A., Hoffman, M.~D., Lee, D., Goodrich, B., Betancourt,
  M., Brubaker, M., Guo, J., Li, P., and Riddell, A. (2017).
\newblock Stan: {{A Probabilistic Programming Language}}.
\newblock {\em Journal of Statistical Software}, 76(1):1--32.

\bibitem[Chang et~al., 2018]{chang2018scalable}
Chang, C., Kundu, S., and Long, Q. (2018).
\newblock Scalable {{Bayesian}} variable selection for structured
  high-dimensional data.
\newblock {\em Biometrics}, 74(4):1372--1382.

\bibitem[Corliss et~al., 2008]{corliss2008sexual}
Corliss, H.~L., Rosario, M., Wypij, D., Fisher, L.~B., and Austin, S.~B.
  (2008).
\newblock Sexual {{Orientation Disparities}} in {{Longitudinal Alcohol Use
  Patterns Among Adolescents}}: {{Findings From}} the {{Growing Up Today
  Study}}.
\newblock {\em Archives of Pediatrics \& Adolescent Medicine},
  162(11):1071--1078.

\bibitem[Du et~al., 2021]{du2021unified}
Du, M., Zhao, H., and Sun, J. (2021).
\newblock A unified approach to variable selection for {{Cox}}'s proportional
  hazards model with interval-censored failure time data.
\newblock {\em Statistical Methods in Medical Research}, 30(8):1833--1849.

\bibitem[Fan and Li, 2001]{fan2001variable}
Fan, J. and Li, R. (2001).
\newblock Variable selection via nonconcave penalized likelihood and its oracle
  properties.
\newblock {\em Journal of the American Statistical Association},
  96(456):1348--1360.

\bibitem[Gelman et~al., 2014]{gelman2014bayesian}
Gelman, A., Carlin, J.~B., Stern, H.~S., Dunson, D.~B., Vehtari, A., and Rubin,
  D.~B. (2014).
\newblock {\em Bayesian Data Analysis}.

\bibitem[Guo, 1993]{guo1993eventhistory}
Guo, G. (1993).
\newblock Event-history analysis for left-truncated data.
\newblock {\em Sociological Methodology}, 23:217--243.

\bibitem[Hern{\'a}n, 2010]{hernan2010hazards}
Hern{\'a}n, M.~A. (2010).
\newblock The hazards of hazard ratios.
\newblock {\em Epidemiology}, 21(1):13--15.

\bibitem[Huang et~al., 2019]{huang2019bilevel}
Huang, H., Shangguan, J., Ruan, P., and Liang, H. (2019).
\newblock Bi-level feature selection in high dimensional {{AFT}} models with
  applications to a genomic study.
\newblock {\em Statistical Applications in Genetics and Molecular Biology},
  18(5).

\bibitem[Kawas et~al., 2000]{kawas2000agespecific}
Kawas, C., Gray, S., Brookmeyer, R., Fozard, J., and Zonderman, A. (2000).
\newblock Age-specific incidence rates of {{Alzheimer}}'s disease: {{The
  Baltimore Longitudinal Study}} of {{Aging}}.
\newblock {\em Neurology}, 54(11):2072--2077.

\bibitem[Konrath et~al., 2015]{konrath2015bayesian}
Konrath, S., Fahrmeir, L., and Kneib, T. (2015).
\newblock Bayesian accelerated failure time models based on penalized mixtures
  of {{Gaussians}}: Regularization and variable selection.
\newblock {\em AStA Advances in Statistical Analysis}, 99(3):259--280.

\bibitem[Kukull et~al., 2002]{kukull2002dementia}
Kukull, W.~A., Higdon, R., Bowen, J.~D., McCormick, W.~C., Teri, L.,
  Schellenberg, G.~D., {van Belle}, G., Jolley, L., and Larson, E.~B. (2002).
\newblock Dementia and {{Alzheimer}} disease incidence: A prospective cohort
  study.
\newblock {\em Archives of Neurology}, 59(11):1737.

\bibitem[Kyung et~al., 2010]{kyung2010penalized}
Kyung, M., Gill, J., Ghosh, M., and Casella, G. (2010).
\newblock Penalized regression, standard errors, and {{Bayesian}} lassos.
\newblock {\em Bayesian Analysis}, 5(2):369--411.

\bibitem[Lee et~al., 2017a]{lee2017variable}
Lee, K.~H., Chakraborty, S., and Sun, J. (2017a).
\newblock Variable selection for high-dimensional genomic data with censored
  outcomes using group lasso prior.
\newblock {\em Computational Statistics \& Data Analysis}, 112:1--13.

\bibitem[Lee et~al., 2017b]{lee2017accelerated}
Lee, K.~H., Rondeau, V., and Haneuse, S. (2017b).
\newblock Accelerated failure time models for semi-competing risks data in the
  presence of complex censoring.
\newblock {\em Biometrics}, 73(4):1401--1412.

\bibitem[Li et~al., 2020]{li2020adaptive}
Li, C., Pak, D., and Todem, D. (2020).
\newblock Adaptive lasso for the {{Cox}} regression with interval censored and
  possibly left truncated data.
\newblock {\em Statistical Methods in Medical Research}, 29(4):1243--1255.

\bibitem[Li and Pati, 2017]{li2017variable}
Li, H. and Pati, D. (2017).
\newblock Variable selection using shrinkage priors.
\newblock {\em Computational Statistics \& Data Analysis}, 107:107--119.

\bibitem[Li and Lin, 2010]{li2010bayesian}
Li, Q. and Lin, N. (2010).
\newblock The {{Bayesian}} elastic net.
\newblock {\em Bayesian Analysis}, 5(1):151--170.

\bibitem[McGough et~al., 2021]{mcgough2021penalized}
McGough, S.~F., Incerti, D., Lyalina, S., Copping, R., Narasimhan, B., and
  Tibshirani, R. (2021).
\newblock Penalized regression for left-truncated and right-censored survival
  data.
\newblock {\em Statistics in Medicine}, 40(25):5487--5500.

\bibitem[Neal, 2011]{neal2011mcmc}
Neal, R.~M. (2011).
\newblock {{MCMC Using Hamiltonian Dynamics}}.
\newblock In Brooks, S., Gelman, A., Jones, G., and Meng, X.-L., editors, {\em
  Handbook of {{Markov Chain Monte Carlo}}}, pages 113--162. {Chapman and
  Hall/CRC}.

\bibitem[Newcombe et~al., 2017]{newcombe2017weibull}
Newcombe, P., Raza~Ali, H., Blows, F., Provenzano, E., Pharoah, P., Caldas, C.,
  and Richardson, S. (2017).
\newblock Weibull regression with {{Bayesian}} variable selection to identify
  prognostic tumour markers of breast cancer survival.
\newblock {\em Statistical Methods in Medical Research}, 26(1):414--436.

\bibitem[Park and Casella, 2008]{park2008bayesian}
Park, T. and Casella, G. (2008).
\newblock The {{Bayesian}} lasso.
\newblock {\em Journal of the American Statistical Association},
  103(482):681--686.

\bibitem[Raman et~al., 2009]{raman2009bayesian}
Raman, S., Fuchs, T.~J., Wild, P.~J., Dahl, E., and Roth, V. (2009).
\newblock The {{Bayesian}} group-{{Lasso}} for analyzing contingency tables.
\newblock In {\em Proceedings of the 26th {{Annual International Conference}}
  on {{Machine Learning}}}, {{ICML}} '09, pages 881--888, {New York, NY, USA}.
  {Association for Computing Machinery}.

\bibitem[Reeder et~al., 2023]{reeder2023characterizing}
Reeder, H.~T., Ha~Lee, K., and Haneuse, S. (2023).
\newblock Characterizing quantile-varying covariate effects under the
  accelerated failure time model.
\newblock {\em Biostatistics}, page kxac052.

\bibitem[Simon and Tibshirani, 2012]{simon2012standardization}
Simon, N. and Tibshirani, R. (2012).
\newblock Standardization and the group lasso penalty.
\newblock {\em Statistica Sinica}, 22(3).

\bibitem[Strittmatter and Roses, 1996]{strittmatter1996apolipoprotein}
Strittmatter, W.~J. and Roses, A.~D. (1996).
\newblock Apolipoprotein {{E}} and {{Alzheimer}}'s {{Disease}}.
\newblock {\em Annual Review of Neuroscience}, 19(1):53--77.

\bibitem[Tanner and Wong, 1987]{tanner1987calculation}
Tanner, M.~A. and Wong, W.~H. (1987).
\newblock The calculation of posterior distributions by data augmentation.
\newblock {\em Journal of the American Statistical Association},
  82(398):528--540.

\bibitem[Tibshirani, 1996]{tibshirani1996regression}
Tibshirani, R. (1996).
\newblock Regression shrinkage and selection via the lasso.
\newblock {\em Journal of the Royal Statistical Society: Series B (Statistical
  Methodology)}, 58(1):267--288.

\bibitem[Tibshirani et~al., 2005]{tibshirani2005sparsity}
Tibshirani, R., Saunders, M., Rosset, S., Zhu, J., and Knight, K. (2005).
\newblock Sparsity and smoothness via the fused lasso.
\newblock {\em Journal of the Royal Statistical Society: Series B (Statistical
  Methodology)}, 67(1):91--108.

\bibitem[Tibshirani, 2009]{tibshirani2009univariate}
Tibshirani, R.~J. (2009).
\newblock Univariate shrinkage in the {{Cox}} model for high dimensional data.
\newblock {\em Statistical Applications in Genetics and Molecular Biology},
  8(1):1--18.

\bibitem[Uno et~al., 2015]{uno2015alternatives}
Uno, H., Wittes, J., Fu, H., Solomon, S.~D., Claggett, B., Tian, L., Cai, T.,
  Pfeffer, M.~A., Evans, S.~R., and Wei, L.-J. (2015).
\newblock Alternatives to hazard ratios for comparing the efficacy or safety of
  therapies in noninferiority studies.
\newblock {\em Annals of Internal Medicine}, 163(2):127--134.

\bibitem[Wilson et~al., 2002]{wilson2002depressive}
Wilson, R.~S., Barnes, L.~L., Mendes De~Leon, C.~F., Aggarwal, N.~T.,
  Schneider, J.~S., Bach, J., Pilat, J., Beckett, L.~A., Arnold, S.~E., Evans,
  D.~A., and Bennett, D.~A. (2002).
\newblock Depressive symptoms, cognitive decline, and risk of {{AD}} in older
  persons.
\newblock {\em Neurology}, 59(3):364--370.

\bibitem[Yuan and Lin, 2006]{yuan2006model}
Yuan, M. and Lin, Y. (2006).
\newblock Model selection and estimation in regression with grouped variables.
\newblock {\em Journal of the Royal Statistical Society: Series B (Statistical
  Methodology)}, 68(1):49--67.

\bibitem[Zhang et~al., 2018]{zhang2018bayesian}
Zhang, Z., Sinha, S., Maiti, T., and Shipp, E. (2018).
\newblock Bayesian variable selection in the accelerated failure time model
  with an application to the surveillance, epidemiology, and end results breast
  cancer data.
\newblock {\em Statistical Methods in Medical Research}, 27(4):971--990.

\bibitem[Zhao et~al., 2020]{zhao2020simultaneous}
Zhao, H., Wu, Q., Li, G., and Sun, J. (2020).
\newblock Simultaneous {{Estimation}} and {{Variable Selection}} for
  {{Interval-Censored Data With Broken Adaptive Ridge Regression}}.
\newblock {\em Journal of the American Statistical Association},
  115(529):204--216.

\bibitem[Zhu et~al., 2019]{zhu2019bayesian}
Zhu, L., Huo, Z., Ma, T., Oesterreich, S., and Tseng, G.~C. (2019).
\newblock Bayesian indicator variable selection to incorporate hierarchical
  overlapping group structure in multi-omics applications.
\newblock {\em The Annals of Applied Statistics}, 13(4).

\end{thebibliography}
